\documentclass[reviewcopy]{elsarticle}

\usepackage[reviewcopy]{adndt}
\usepackage{longtable}

\usepackage{mathptmx}

\usepackage{amsmath}

\usepackage{amssymb}

\biboptions{square,sort&compress}
\bibpunct[]{[}{]}{,}{n}{}{;}
\begin{document}

\begin{frontmatter}

\journal{Atomic Data and Nuclear Data Tables}


\title{Energy levels and radiative rates for transitions in Fe~V, Co~VI and Ni~VII}

  \author[One]{K. M. Aggarwal\corref{cor1}}
  \ead{K.Aggarwal@qub.ac.uk}
  \author[Two]{P. Bogdanovich\fnref{}} 
  \author[One]{F. P. Keenan\fnref{}}
  \author[Two]{R. Kisielius\fnref{}} 


  \cortext[cor1]{Corresponding author.}

  \address[One]{Astrophysics Research Centre, School of Mathematics and Physics, Queen's University Belfast,\\Belfast BT7 1NN,
Northern Ireland, UK}

  \address[Two]{Institute of Theoretical Physics and Astronomy, Vilnius University, Saul{\.e}tekio al. 3, LT-10222 Vilnius, Lithuania}

\date{06/05/2016} 

\begin{abstract}  
Energy levels, Land\'{e} $g$-factors and radiative lifetimes are reported for 
the lowest 182 levels of the 3d$^4$, 3d$^3$4s and 3d$^3$4p configurations of 
Fe~V, Co~VI and Ni~VII. Additionally, radiative rates ($A$-values) have 
been calculated for the E1, E2 and M1 transitions among these levels.  
The calculations have been performed in a quasi-relativistic  approach (QR) 
with a very large {\em configuration interaction} (CI) wavefunction expansion, 
which has been found to be necessary for these ions. Our calculated energies for all ions 
are in excellent agreement with the available measurements, 
for most levels. Discrepancies among various calculations for the radiative rates 
of E1 transitions in Fe~V are up to a factor of two for stronger transitions 
($f \geq 0.1$), and larger (over an order of magnitude) for weaker ones. The 
reasons for these discrepancies have been discussed and mainly are due to
the differing amount of CI and methodologies adopted. However, there are no 
appreciable discrepancies in similar data for M1 and E2 transitions, or the 
$g$-factors for the levels of Fe~V, the only ion for which comparisons are 
feasible. 
 \\

Received 13 May 2016; Accepted 26 May 2016 \\


\end{abstract}

\end{frontmatter}




\newpage

\tableofcontents
\listofDtables
\listofDfigures
\vskip5pc


\section{Introduction}
\label{intro}
Emission lines of iron group elements, particularly of Fe and Ni,  show
rich spectra covering a wide wavelength range in a variety of solar 
and astrophysical plasmas. Their lines are observed from almost all ionisation 
stages as may be noted from the {\em Atomic Line List (v2.04)} of Peter van Hoof 
(${\tt {\verb+http://www.pa.uky.edu/~peter/atomic/+}}$), CHIANTI database 
\cite{chianti1,chianti8} at ${\tt {\verb+http://www.chiantidatabase.org+}}$ 
and the atomic and molecular database Stout \cite{stout}. Similarly, many of 
these elements are also useful for studies of fusion plasmas. However, to reliably
model the spectral lines in  plasmas, atomic data are required for several parameters, such as  
energy levels and radiative rates ($A$-values). Therefore, over the past few 
decades several workers have reported data for many such ions, including 
ourselves -- see for example \cite{niions, fe15, rk1}. However, (in general) 
most of the work has been performed for highly ionised systems and comparatively 
less attention has been paid to the lowly ionised species. This is because 
such ions are more problematic  and usually require much larger calculations 
to achieve a reasonably satisfactory level of accuracy.

Iron is a very important element for both astrophysical and fusion plasma 
studies, and emission and absorption lines of Fe~V have been observed in many 
hot stars and nebulae -- see for example, Kramida \cite{ak} and references 
therein. Its lines have also been observed in white dwarfs \cite{dwarf} and 
are  useful for the study of the fine-structure constant in a gravitational 
field. The first investigation of the Fe~V spectrum was undertaken 
as early as 1937  by Bowen \cite{isb1}, who identified 57 levels of the 3d$^4$, 3d$^3$4s and 
3d$^3$4p configurations. This  study was subsequently extended  by other workers, 
such as \cite{bcf, ek}. Therefore, a very rich experimental spectrum of high accuracy, 
involving as many as 982 lines, is available for this ion \cite{ek}. 
A critical compilation of all measured lines of several ions with 19 $\le$ Z $\le$ 28 
 was undertaken by Sugar and Corliss \cite{sug}, 
and their recommended energy levels are also available on the NIST (National 
Institute of Standards and Technology) website 
{\tt http://www.nist.gov/pml/data/asd.cfm} \cite{nist15}. Later, Azarov 
et al. \cite{aza}  also measured many lines of the 3d$^3$4d and 3d$^3$5s 
configurations of Fe~V. A similar situation exists for Co~VI \cite{sug}, 
and as for Fe~V, its lines were  studied as early as 1938 \cite{isb2, pk}. 
However, the observed spectrum of Ni~VII is not as rich as for the other 
Ti-like ions Fe~V and Co~VI, because many levels are missing for the 3d$^4$ and 3d$^3$4p 
configurations and none has been measured for  3d$^3$4s -- see 
Table~\ref{ene-ni7} or the 
NIST website. Additionally, the situation regarding  radiative data 
($A$-values) is even worse, particularly for Co~VI and Ni~VII, although some 
results are available for Fe~V  \cite{aza, jrf, sn, np, om}. Therefore, 
in this paper we  calculate energy levels and $A$-values for 
three Ti-like ions, namely Fe~V, Co~VI and Ni~VII.

As noted above, calculations for lowly ionised  ions are generally not 
 straightforward, and hence require a significant amount of effort. This also applies to Ti-like species. 
 Early  calculations for energy levels were performed by Ekberg \cite{ek}, who adopted 
a least-square fit to the observed values, apart from applying a few 
corrections. In spite of this, differences between the observed and calculated 
energies are between $+$299 and $-$470 cm$^{-1}$ (see tables III--V of 
\cite{ek}), although they equate to less than 0.2\%. Later, O`Malley et al. 
\cite{om} performed relativistic configuration interaction (RCI) calculations 
with $\sim 15\,000$ vectors, and determined energies for 5 ($J = 0$) levels of  the
3d$^4$ and 19 ($J = 1$) of 3d$^3$4p configurations. They achieved a good 
accuracy within $\sim 3\%$ of the measurements -- see their table~III. The 
largest {\em ab initio} calculation available so far is by Nahar and Pradhan 
\cite{np}, who adopted the Breit-Pauli $R$-matrix method to calculate energies 
for 3865 levels of Fe~V. However, the main problem with their work is that differences 
with measurements are up to 10\%, for several levels and of all configurations 
-- see their table~III or table~III of \cite{om} for a shorter version. The most 
difficult to determine are  the energy levels of the 3d$^4$ configuration, as 
may also be noted from table~1 of Ballance et al. \cite{bgm}, who adopted the 
general-purpose relativistic atomic structure package ({\sc grasp}) to calculate 
energies for 359 levels of the 3d$^4$, 3d$^3$4s, 3d$^3$4p, 3d$^3$4d, and 
3d$2$4s$^2$ configurations. Since their focus was on the calculation of 
collisional data, they could only include a limited CI (configuration 
interaction), but differences between their energies and those of NIST are up 
to $16\%$ for several levels, particularly those belonging to 3d$^4$.

Adopting the same {\sc grasp} code as by \cite{bgm}, we have performed our 
calculations with much more extensive CI,  but differences with the 
NIST compilation remain significant, both in magnitude and orderings, 
particularly for the lowest 34 levels of the 3d$^4$ configuration. Therefore, we 
employed the flexible atomic code ({\sc fac}) of Gu~\cite{gu} which (generally) 
provides results of comparable accuracy with other atomic structure codes, but 
 is much more efficient to run 
and hence saves both computational and human time. Unfortunately, the 
results obtained with this code are as unsatisfactory as with {\sc grasp}. To be 
specific, we included CI with up to 100\,915 levels ($n \leq 5$), but 
differences in energy for the levels of the 3d$^4$ configuration of Fe~V are up 
to 15\%, as shown in Table~\ref{table_A}. Therefore, it became clear that we either 
have to extend the CI to a much higher order, or have to apply another approach, 
such as the use of non-orthogonal orbitals. However, having recently gained experience 
from our work on Cr-like ions \cite{rk1, rk2}, we have employed the 
quasi-relativistic approximation (QR) \cite{tro08}. 

\section{Details of calculations}
\label{calc}

In this work we investigate the lowest two even-parity configurations 3d$^4$ 
and 3d$^3$4s with 72 energy levels and one odd-parity configuration 3d$^3$4p
with 110 levels. We utilise the quasi-relativistic (QR) approach \cite{tro08} 
 as it was done in our previous studies \cite{rk1,rk2} of 
spectroscopic parameters for  iron peak elements. At the start of the
calculations we solve quasi-relativistic Hartree-Fock equations (QRHF)
\cite{qr06} for the ground configuration, and determine all one-electron
radial orbitals (RO) for electrons with principal quantum number 
$N \leq 3$. Next  we solve QRHF equations in the frozen-core
potential for all $4\ell$ electrons ($\ell \leq 3$) for the configurations
3d$^3$4$\ell$. Subsequently the determined RO basis is extended  by including 
transformed radial orbitals (TRO) \cite{tro08}  to effectively account 
for correlation effects \cite{bk99}. TROs are determined for electrons 
with principal quantum number $5 \leq n \leq 10$ and all allowed values
of the orbital quantum number $\ell < n$. Using this methodology, our basis
consists of 55 ROs. The same ROs are utilised both for even and odd 
configurations. This way we avoid issues with RO non-orthogonality, 
important in the calculation of radiative transitions. Inaccuracies in level
energies arising from that approximation are minimised by the adoption of a
large CI basis.

The correlation effects are included using the CI method. Therefore a list of
admixed configurations (AC) is constructed for each investigated 
configuration (adjusted configuration). This AC list is composed by including one- 
and two-electron promotions from the active shells ($3\ell$ and $4\ell$) of the
investigated configuration to all those of the same parity, which can
be described by the RO basis generated earlier. The presence of various admixed
configurations in the CI basis dictates what kind of additional symmetries are
included in the eigen-functions of the investigated configurations. Thus the
number of ACs can be considered as the main criterion for the inclusion of 
electron correlation effects.

The adopted RO basis includes one-electron radial orbitals having orbital
momenta from $\ell = 0$ to $\ell = 9$. Their combinations in the admixed 
configurations enable us  to construct nearly all necessary symmetries of momenta. 
Therefore the method of TRO construction \cite{tro08, bk99} and extensive set of
the principal quantum numbers $n$ ensures a very effective inclusion of the radial
correlations.

Parameters of the calculation for the Ti-like ions under consideration are 
presented in Table~\ref{table_B}. The large maximum numbers of AC for the even
$M^{\rm e}_{\rm AC}$ and odd $M^{\rm o}_{\rm AC}$ configurations, together with
possible configuration state function (CSF) numbers ($M^{\rm e}_{\rm CSF}$ and 
$M^{\rm o}_{\rm CSF}$) given in this table, indicate that it is impossible to 
include into the CI wavefunction expansion all CSFs originating from the arranged 
AC sets. Therefore one needs to select the admixed configurations according to 
their average contributions into the eigen-functions of the investigated 
configurations. The contributions are determined in the second order of the 
perturbation theory -- see \cite{bkr07}. We apply the selection criteria 
$w = 10^{-6}$, i.e. all  ACs with the average contribution ${\bar w} < w$ are excluded 
from the calculations.

The two even-parity configurations,  namely 3d$^3$4d and 
3d$^2$4s$^2$, are close to the investigated configurations in their energies, 
and hence strongly affect these. To correctly determine their influence
and to account more consistently for the 3- and 4-electron correlation
corrections, the set of selected ACs is extended by adding the admixed 
configurations that interact strongly with above mentioned (3d$^3$4d and 
3d$^2$4s$^2$) configurations. The selection criteria  for these configurations
is much larger ($w=10^{-3}$). In the case of odd-parity configurations, the 
additional admixed configurations are generated for the  
3s$^2$3p$^5$3d$^5$, 3s$^2$3p$^6$3d$^3$4f, 3s$^2$3p$^6$3d$^2$4s4p, and 3s$^2$3p$^6$3d4s$^2$4p
set of AC. The numbers $S$ in Table~\ref{table_B} represent the reduced (even and odd) 
configurations included in the CI basis, which are  about  4 to
5 times smaller than the initial ones. 

A comparison of $S$ values for the three ions considered here demonstrates
that, for the same configuration selection criteria $w$, the number of selected 
configurations (slightly) decreases as the degree of ionisation increases. Such  behaviour 
confirms the well-known fact that the importance of correlation effects decreases
with increase of the electrostatic potential affecting moving electrons.

While performing actual CI calculations, the value of the $S$ parameter is not so 
important compared to the number of CSF ($C$) generated by the configurations
included in the CI basis. Corresponding C-values  for the even
 and   odd configurations are also given in  
Table~\ref{table_B}. We note that their values are quite large (e.g.
$C^{\rm o} \sim 10^7$), and it becomes time consuming to perform 
calculations for Hamiltonian matrices of such sizes. 

At the next step we  reduce the number of CSF, a
 procedure which relies on the relocation of the virtually excited electrons
to the front of the active shells of AC. We further discard those CSFs 
which have off-diagonal matrix elements of operator, describing electrostatic 
interaction with the investigated configurations, equal to zero \cite{csf02}.
The numbers of CSF after these reductions are given as $R^{\rm e}$ and $R^{\rm o}$
in Table~\ref{table_B}. One can see that this step reduces the basis of CSFs by
almost an order of magnitude. We note that this type of  significant CSF reduction 
does not affect the effectiveness of the CI wavefunction expansion. Interestingly,  while 
the ionisation degree increases and consequently the number of selected 
configurations $S$ decreases, the number $R$ of produced CSFs  increases.
This behaviour demonstrates that the above described AC reduction procedure leads to
the inclusion of different configurations for different degrees of  ionisation in 
the isoelectronic sequence. Therefore the values of $R$ can increase.

In our computational method, the most important factor limiting the calculation
is the number of CSFs with the same total $LS$ momenta. For the 
 Fe~V, Co~VI, and Ni~VII Ti-like ions considered here, the largest number $T$ of same $LS$ 
momenta is attributed to the $^3$F term, both for even and odd configurations, given in 
Table~\ref{table_B}. It is clear that their values are proportional to 
$R^{\rm e}$ and $R^{\rm o}$.

Correlation effects are very important  for medium ionisation stage
ions with an open 3d shell. When we implement the CI model, we include a huge
 number of admixed interacting configurations, but our  limited   computing
resources necessitate  some compromises -- see Table~\ref{table_B}. 
Although each separate configuration cannot significantly  affect the calculated 
results, the combined influence of such (omitted) configurations is comparatively 
appreciable, and hence causes some discrepancies between the
calculated and experimental level energies. Therefore, we reduce integrals of
the electrostatic interaction for all investigated configurations by $1.3\%$, as 
 in \cite{rk2}. Such a minimal change of integral values noticeably reduces 
discrepancies in the theoretical level energies, leading to more accurate transition 
wavelengths. This in turn reduces the influence of errors in transition 
energies, and subsequently  on transition parameters.

Relativistic effects are included in the Breit-Pauli approximation as
described in \cite{tro08}. The level energies of the investigated configurations
and their eigen-functions are determined by diagonalising the Hamiltonian matrix.
These data are utilised to determine radiative transition parameters for  
electric dipole (E1), electric octupole (E3), and magnetic dipole (M2) 
transitions among the levels of even- and odd-parity configurations, and for
 magnetic dipole (M1) and electric quadrupole (E2) transitions among the
levels of the same parity configurations -- see section \ref{rad}. These parameters  
 are further used to determine the total radiative lifetimes $\tau$ of the
excited levels. By utilising the determined CI wavefunctions, we also compute
electron-impact excitation cross sections and rates in the plane-wave
Born approximation. These parameters are not discussed in the present paper but
they are freely available from the database ADAMANT 
({\tt http://www.adamant.tfai.vu.lt/database}). 

Apart from our own computer codes developed specifically for the calculation 
of spectroscopic parameters and electron-impact excitation cross-sections in the
QR approximation, we adapt the codes from the MCHF package \cite{ah91, cff91, 
cffmg91} for use of the quasi-relativistic radial orbitals.

\section{Energy levels and Land\'{e} $g$-factors}
\label{ene}

Level energies obtained in the QR approximation are listed in 
Table~\ref{table_A} for all 34 levels of the 3d$^4$ configuration of 
Fe~V, and  agreement with the corresponding experimental data of NIST is 
highly satisfactory. The ordering of the levels is also the same in both theory 
and measurements. Generally, our calculated energies are slightly higher, but 
the discrepancies for most of the levels are less than $1.0\%$, except for  seven which
 deviate by up to $1.24\%$. The largest relative discrepancy 
of $1.48\%$ is for  level 23 ($\,^1_4{\rm S}_0$). On the other hand, the 
highest level $\,^1_0{\rm S}_0$ of the ground configuration 3d$^4$ shows
the largest absolute discrepancy of $818~{\rm cm}^{-1}$
($0.83\%$). The averaged relative disagreement for the levels of the 3d$^4$ configuration is only
$0.83\%$. More importantly, agreement between our calculations and the NIST 
compilations is much better (within 0.5\%) for levels of the 3d$^3$4s and 
3d$^3$4p configurations -- see Table~\ref{ene-fe5} in which energies for all 
182 levels of Fe~V are listed. The averaged relative discrepancy for the excited
configuration levels is only $0.16\%$, and is $0.12\%$ for levels of the
even-parity configuration 3d$^3$4s and  $0.17\%$ for the odd-parity 
 3d$^3$4p.  This good agreement  for 
a larger number of levels is highly satisfactory and encouraging. However, we 
note  that the $LSJ$ designations listed in the table are not always 
definitive, because we have performed just a formal identification based on the 
maximum percentage contribution of a particular CSF in the CI wavefunction 
expansion, and some levels are highly affected by CSF mixing. For this reason 
their description using just a simple $LSJ$ notation is not definitive in all 
cases, and other, more sophisticated level identification schemes have to be 
applied instead of an $LS$ designation. All such levels are shown by a 
superscript ``a"  -- see e.g., levels 83, 87, 89, and 104 in 
Table~\ref{ene-fe5}. However, this is a rather general atomic structure problem, 
as also noted in our earlier papers \cite{rk1, rk2}.

In Table~\ref{ene-co6} we compare our calculated energies with the NIST 
compilations for all 182 levels of Co~VI. As for Fe~V,  measurements 
are  available for most levels, and  discrepancies with these are 
slightly lower. The averaged relative discrepancy for the ground configuration
is $0.75\%$. Similar to  Fe~V, the largest relative disagreement is for
 level $\,^1_4{\rm S}_0$. The averaged relative discrepancy for the excited
configurations is only $0.12\%$, with $0.16\%$ for 
the 3d$^3$4s configuration and only  $0.048\%$ for 3d$^3$4p. 

Unfortunately, as it has been stated in Section~\ref{intro}, the number of 
levels for which measurements are available is very limited for Ni~VII. 
Therefore it is not used to calculate and compare the averaged relative 
discrepancies. Nevertheless, in Table~\ref{ene-ni7} we list our calculated energies 
for all the 182 levels of Ni~VII along with those of NIST. 
The differences between the theoretical and experimental energies are smaller that
$0.8\%$, excluding level 2 where it is 1.4\%
($4~ {\rm cm}^{-1}$). The discrepancies are no greater than $301~{\rm cm}^{-1}$, and below 
$0.1\%$ for common levels of the 3d$^3$4p configuration. Therefore, for all 
three Ti-like ions Fe~V, Co~VI and Ni~VII there are no significant 
discrepancies for energy levels between theory and measurements, and therefore 
our results listed in Tables~\ref{ene-fe5},\ref{ene-co6},\ref{ene-ni7} can be 
confidently applied to the modelling of plasmas. 

For all three  ions investigated the QR calculations are performed in the same 
approximation. Consequently, a comparison of the discrepancies for specific
 level energies in  Fe~V and Co~VI  enables us to draw  
conclusions about the accuracy of the theoretical energies for those Ni~VII levels 
which have no  experimental data.

Finally, we note that data in the Tables~\ref{ene-fe5},\ref{ene-co6},\ref{ene-ni7}  are provided for only the lowest 
182 levels of the 3d$^4$, 3d$^3$4s and 3d$^3$4p configurations. Inclusion of 
similar results for  levels of the 3d$^3$4d or 3d$^3$4f configurations is 
not feasible, because these cover a much wider energy range (and number over 1000) 
and intermix with  many levels from other configurations (such as 3p$^5$3d$^5$ and 
3d$^3$5$\ell$), whereas there is no such intermixing among 
the lowest 182.

Also listed in Tables~\ref{ene-fe5},\ref{ene-co6},\ref{ene-ni7} are the
 Land\'{e} $g$-factors (dimensionless) that show the splitting of energy levels 
in a magnetic field, and represent the Zeeman effect for a particular $LSJ$ 
level. It is given by
\begin{equation}
g = 1 + \sum_{CLS}{\alpha(CLSJ) \frac{J(J+1)-L(L+1)+S(S+1)}{2J(J+1)}}
\end{equation}
where the sum is over all CSFs for that level, $C$ is the configuration, $LSJ$ 
are total moments of the level, and $\alpha(CLSJ)$ is a weight (a square of the
expansion coefficient) of a particular CSF for the level eigen-function. Sometimes 
measurement of $g$ are available and hence may help in assessing the accuracy 
of the calculations. Unfortunately, for the ions  considered here no 
experimental results are available with which to compare our data, but O`Malley 
et al. \cite{om} have reported $g$-factors for 19 ($J = 1$) levels of the 
3d$^3$4p configuration of Fe~V calculated in the relativistic configuration
interaction (RCI) approximation. Therefore, in Table~\ref{table_C} we have 
included their and our $g$-factors for  ready comparison. For most levels 
there are no discrepancies between the two independent calculations, but our 
results are lower by $\sim 40\%$ for two, namely 89 ($^5$F$^{\rm o}_1$) and 138 
($^3$D$^{\rm o}_1$). The $g$-factors are sensitive to primarily those levels which 
have low $LS$-purity, and hence the differences between the two calculations.

\section{Radiative rates and lifetimes}
\label{rad}
Apart from spectral modelling (including diagnostics) and the determination of 
the total radiative lifetimes ($\tau$),\,\,  $A$-values are required for 
calculations of local thermodynamic equilibrium (LTE) in stellar opacities, 
and radiative levitation and acceleration of heavy elements -- see for example, 
\cite{np} and references therein. For this reason, Nahar and Pradhan \cite{np} 
performed very large calculations of energy levels and E1\, $A$-values for 
transitions in Fe~V, as already stated in Section~\ref{intro}. 
However, for more sophisticated modelling applications, and particularly the 
determination of $\tau$, corresponding results for the electric quadrupole E2, 
magnetic dipole M1, and magnetic quadrupole M2 transitions are also desirable. 
Therefore, in a separate paper \cite{sn} they reported $A$-values for the M1 
and E2 transitions of Fe V.

In Tables~\ref{tran-fe5},\ref{tran-co6},\ref{tran-ni7} we list 
transition energies ($\Delta E$, cm$^{-1}$), wavelengths ($\lambda$, 
\AA), emission radiative rates ($A$-values, s$^{-1}$), weighted oscillator 
strengths ($gf$, dimensionless), and transition line strengths ($S$-values in 
atomic units) for the E1, E2 and M1  transitions of Fe~V, Cu~VI and Ni~VII, 
respectively. These results are among the 182 levels listed in 
Tables~\ref{ene-fe5},\ref{ene-co6},\ref{ene-ni7}, but we only include those transitions with $A$-values (and 
other parameters)  which are $\geq 10\%$ of the largest value for 
an emission transition probability from the upper level $j$. Hence to save 
on space data for  very weak transitions are not provided, as their impact on plasma
modelling  should be insignificant. For the same reason,  \,$A$-values 
for the M2 and E3 transitions are also not included in 
Tables~\ref{tran-fe5},\ref{tran-co6},\ref{tran-ni7}. However, \,$A$-values (and other related 
parameters) for all (including much weaker) transitions, along with 
electron-impact excitation data determined in the plane-wave Born approximation, 
are freely available in  ASCII format from the  ADAMANT database at Vilnius 
University ({\tt http://www.adamant.tfai.vu.lt/database}). 

Additionally, we list $\lambda$\,(\AA) and $f$-values (dimensionless) for all 
absorption E1 transitions with $f \geq 0.1$ in 
Tables~\ref{abs-fe5},\ref{abs-co6},\ref{abs-ni7}. This is because not all important 
absorption transitions are present in Tables~\ref{tran-fe5},\ref{tran-co6},\ref{tran-ni7} 
(due to selection rules), and hence these results may be 
helpful for future comparisons and  accuracy assessments. 
Also listed in these Tables are the $\lambda$ and $f$-values for some weaker 
($f \geq 0.001$) absorption lines originating from the lowest 5 levels of the 
ground configuration term 3d$^4$\,$^5D$. These lines may have applications in the  
modelling of the absorption spectra of low-temperature plasmas.

$A$-values for E1 transitions of Fe~V are available in the literature, mainly 
by \cite{np, om}. Additionally, Garstang \cite{ gar} has reported $A$-values 
for the M1 and E2 transitions, but only  among  levels of the lowest 3d$^4$ 
configuration. In Table~\ref{table_D} we compare our results for some E1 
transitions of Fe~V with those of \cite{bcf,np,om}. In general, the $f$-values 
of Fawcett \cite{bcf} and O`Malley et al. \cite{om} show good agreement with our results, 
although differences for a few are up to a factor of two, which include some 
(comparatively) strong transitions, such as $23 - 132$ and $34 - 182$. 
Similarly, our data agree closely  with those of \cite{om}, particularly for 
strong transitions, although differences are up to a factor of two for some 
weaker ones, such as: $1-89$, $6-133$ and $23-138$. However, the maximum 
discrepancies for any set of $f$-values listed in Table~\ref{table_D} are with 
the BPRM results of Nahar and Pradhan \cite{np}, and this includes both the 
strong ($1-80$ and $23-132$) and weak ($1-82$ and $6-133$) transitions. For 
these (and other) transitions the $f$-values of \cite{np} differ by over an 
order of magnitude with other results. Differences in $f$-values between any two calculations can 
often be large (i.e. a factor of two or more for some strong transitions) as 
seen in Table~\ref{table_D} or in table~VI of \cite{np}. Such differences 
mainly arise with the varying amount of CI adopted in a calculation as well 
as the methodology applied, as discussed and demonstrated earlier by Aggarwal 
et al. \cite{fe15} for three Mg-like ions. However, based on the comparisons 
shown in Table~\ref{table_D} and noting the large discrepancies in the energy 
levels of Nahar and Pradhan \cite{np} in section~\ref{ene}, their radiative 
data appear to be comparatively less accurate.

In Table~\ref{table_E} we compare our $A$-values with those of Garstang 
\cite{gar} for the M1 transitions among the levels of the 3d$^4$ configuration.
These transitions are comparatively stronger than the corresponding E2 ones among 
these levels, also reported in \cite{gar}. Similar results of \cite{sn} for 
these transitions are not included in this table, because there are no 
discrepancies with the data of \cite{gar} -- see table 6 of \cite{sn}.
Considering the low strengths of these transitions, the agreement among three 
independent calculations is highly satisfactory. The only exceptions are the 
$4-7$ and $5-7$ transitions for which the $A$-values of \cite{gar} appear to 
be interchanged. 
For these two transitions (as for others) there are no significant discrepancies 
between our $A$-values ($1.18 \times 10^{-3}$ and $6.24 \times 10^{-3}$
s$^{-1}$) and those from \cite{sn} ($8.34 \times 10^{-4}$ and 
$4.34 \times 10^{-3}$~ s$^{-1}$). Since \cite{sn} have also reported $A$-values 
for the E2 transitions, in Table~\ref{table_F} we show comparisons for a few, 
particularly those with larger strengths. As for M1 transitions, for these E2
also there are no discrepancies between the two calculations, except that there 
is a difference of about a factor of two, and our results are lower. This is because 
there is a difference of a factor of 2/3 in the definitions of $A$-values for 
the E2 transitions -- see Eq.~(4) of \cite{fe17} and Eq.~(11) of \cite{sn}. 
A similar problem was noted earlier for the E2 transitions of Fe~XVII 
\cite{fe17}, and our definitions of $A$-values and transition strengths $S$ 
correspond to those adopted by the NIST.

As for other ions, we have also calculated lifetimes 
($\tau$ = 1.0/$\sum_{i} A_{ji}$, s), where the sum is over all calculated 
radiative decay channels with $i<j$. For the calculations we include $A$-values 
for all E1, E2 and M1 transitions, and list our results in 
Tables~\ref{ene-fe5},\ref{ene-co6},\ref{ene-ni7} for Fe~V, Co~VI and Ni~VII, respectively. The only 
data available in the literature for comparison are by Bi\'{e}mont et al. 
\cite{eb} for the 3d$^4$~$^5$D$_3$ level, which are 374.3, 140.2 and 58.9~s, 
for Fe~V, Co~VI and Ni~VII, respectively, which compare favourably with our corresponding 
values of 379, 138 and 58~s.

\section{Conclusions}

In this work we have reported energy levels, Land\'{e} $g$-factors and the total
radiative lifetimes $\tau$ for the lowest 182 levels of the three Ti-like ions  
Fe~V, Co~VI and Ni~VII. These levels belong to the 3d$^4$, 3d$^3$4s and 
3d$^3$4p configurations, and do not have intermixing with those from others, 
such as 3d$^3$4d and 3d$^3$4f. Experimental energies are available for most 
levels of Fe~V and Co~VI, but for only a few of Ni~VII. 

A large portion of the theoretical level energies differ from the experimental data 
 by only a few hundreds of cm$^{-1}$ or even less. These discrepancies 
decrease as the ionisation degree increases. As a consequence, the averaged 
discrepancies for the ground configuration levels are $0.83\%$ for Fe~V and 
 $0.75\%$ for Co~VI. For the excited configurations 
where the level energies are larger, these disagreements are noticeably smaller and 
decrease to $0.12\%$ for both Fe~V and Co~VI. There is a lack of  experimental
level energies for Ni~VII, but  agreement with our results for levels in
 common is very good. The largest relative discrepancy for the 3d$^3$4p 
configuration is just $0.13\%$, and is less than $0.1\%$ for  most other levels.
This leads to the conclusion that our calculated level energies and the transition
wavelengths for Ni~VII are highly accurate, and hence  suitable for line identifications
in future experiments.

For all three ions the radiative lifetimes $\tau$ and the Land\'{e} 
$g$-factors are consistently determined for the first time. There are no available theoretical 
or experimental   $\tau$ data for comparison purposes,  but there are no
appreciable disagreements with previous theoretical results of $g$, available for
 only a few levels.

Radiative rates for the three ions have also been reported for all E1, E2 and 
M1 emission transitions. Earlier data for the E1 transitions are available for 
Fe~V by \cite{sn,np}. However, in comparison to our calculations and those of 
others \cite{bcf,om}, their $A$-values appear to be less accurate, and so are 
their energy levels which differ from the measurements and our work by some 
$10\%$ for many levels. Unfortunately, no such data are available for 
transitions in Co~VI and Ni~VII. Among other types, \,$A$-values for the M1 and 
E2 transitions are also available \cite{sn,gar}, but only among the levels of 
the 3d$^4$ and 3d$^3$4s configuration of Fe~V. The M1 transitions are 
comparatively stronger than E2, and there are no discrepancies between the 
present and the earlier results for any type of radiative transition. 
However, the present data cover the full range of all types of transitions 
among the lowest 182 levels. We believe our present data  will be useful not 
only for the modelling of plasmas but also for further accuracy assessments. 
\\ \\

\clearpage

\renewcommand{\baselinestretch}{1.5}

\footnotesize  

\begin{longtable}{rlrrr}
\caption{\label{table_A}
Comparison of energy levels (in cm$^{-1}$) of the 3d$^4$ configuration 
of Fe\,V.
}
Index &
\multicolumn{1}{l}{Level} &
\multicolumn{1}{r}{NIST} &
\multicolumn{1}{r}{QR} &
\multicolumn{1}{r}{FAC} \\
\hline
\endfirsthead
\caption[]{(continued)}
Index &
\multicolumn{1}{l}{Level} &
\multicolumn{1}{r}{NIST} &
\multicolumn{1}{r}{QR} &
\multicolumn{1}{r}{FAC} \\
\hline
\endhead
1    & $^5_4$D$_0$ & 0      & 0      &      0 \\
2    & $^5_4$D$_1$ & 142    & 144    &    134 \\
3    & $^5_4$D$_2$ & 418    & 418    &    391 \\
4    & $^5_4$D$_3$ & 803    & 803    &    753 \\
5    & $^5_4$D$_4$ & 1283   & 1280   &   1204 \\
6    & $^3_4$P$_0$ & 24056  & 24315  &  24276 \\
7    & $^3_4$H$_4$ & 24932  & 25134  &  28506 \\
8    & $^3_4$P$_1$ & 24973  & 25238  &  25141 \\
9    & $^3_4$H$_5$ & 25226  & 25420  &  28890 \\
10   & $^3_4$H$_6$ & 25528  & 25715  &  29180 \\
11   & $^3_4$P$_2$ & 26468  & 26748  &  26560 \\
12   & $^3_4$F$_2$ & 26761  & 27036  &  28446 \\
13   & $^3_4$F$_3$ & 26842  & 27110  &  28577 \\
14   & $^3_4$F$_4$ & 26974  & 27234  &  28831 \\
15   & $^3_4$G$_3$ & 29817  & 30095  &  33120 \\
16   & $^3_4$G$_4$ & 30147  & 30419  &  33432 \\
17   & $^3_4$G$_5$ & 30430  & 30686  &  33740 \\
18   & $^3_4$D$_3$ & 36630  & 36985  &  39556 \\
19   & $^1_4$G$_4$ & 36586  & 37041  &  39133 \\
20   & $^3_4$D$_2$ & 36758  & 37123  &  39666 \\
21   & $^3_4$D$_1$ & 36925  & 37296  &  39826 \\
22   & $^1_4$I$_6$ & 37512  & 37822  &  43006 \\
23   & $^1_4$S$_0$ & 39633  & 40221  &  40264 \\
24   & $^1_4$D$_2$ & 46291  & 46651  &  48886 \\
25   & $^1_4$F$_3$ & 52733  & 53173  &  57311 \\
26   & $^3_2$P$_2$ & 61854  & 62275  &  65971 \\
27   & $^3_2$F$_4$ & 62238  & 62642  &  66758 \\
28   & $^3_2$F$_2$ & 62321  & 62798  &  66859 \\
29   & $^3_2$F$_3$ & 62364  & 62812  &  66885 \\
30   & $^3_2$P$_1$ & 62914  & 63366  &  66975 \\
31   & $^3_2$P$_0$ & 63420  & 63890  &  67451 \\
32   & $^1_2$G$_4$ & 71280  & 71773  &  77163 \\
33   & $^1_2$D$_2$ & 93833  & 94559  & 100790 \\
34   & $^1_0$S$_0$ & 121130 & 121948 & 127476 \\
\hline               
\end{longtable}

\begin{flushleft}
NIST: {\tt http://www.nist.gov/pml/data/asd.cfm} \\
QR: Present results in the QR approximation  \\
FAC: Present results with the FAC code with 100\,915 level calculations \\
\end{flushleft}

\clearpage

\begin{longtable}{lrrr}
\caption{\label{table_B}
Number of configurations and CSF adopted in the QR calculations.
}
Parameter &
\multicolumn{1}{c}{Fe\,V} &
\multicolumn{1}{c}{Co\,VI} &
\multicolumn{1}{c}{Ni\,VII} \\
\hline
\endfirsthead\\
\caption[]{(continued)}
Parameter &
\multicolumn{1}{c}{Fe\,V} &
\multicolumn{1}{c}{Co\,VI} &
\multicolumn{1}{c}{Ni\,VII} \\
\hline
\endhead
$M^{\rm e}_{\rm AC}$  & 4536         & 4536         &  4536        \\
$M^{\rm o}_{\rm AC}$  & 3412         & 3412         &  3412        \\
$M^{\rm e}_{\rm CSF}$ & 26\,770\,069 & 26\,770\,069 & 26\,770\,069 \\
$M^{\rm o}_{\rm CSF}$ & 41\,878\,914 & 41\,878,'914 & 41\,878\,914 \\
$S^{\rm e}$           & 1103         & 1076         & 1007         \\
$S^{\rm o}$           & 672          & 656          & 617          \\
$C^{\rm e}$           & 6\,628\,071  & 6\,411\,971  & 5\,802\,821  \\
$C^{\rm o}$           & 9\,739\,792  & 9\,468\,640  & 8\,648\,190  \\
$R^{\rm e}$           & 663\,037     & 643\,672     & 602\,899     \\
$R^{\rm o}$           & 876\,445     & 902\,259     & 903\,614     \\
$T^{\rm e}$           & 86\,177      & 83\,861      & 78\,629      \\
$T^{\rm o}$           & 89\,331      & 91\,810      & 91\,595      \\
\hline   
\end{longtable}



\begin{longtable}{rllll}
\caption{\label{table_C}
Comparison of Land\'{e} $g$-factors (dimensionless) for the 3d$^3$3p ($J = 1$) 
levels of Fe~V. \\See Table~1 for definition of all levels.
}
Index &
\multicolumn{1}{c}{Configuration} &
\multicolumn{1}{c}{Level} &
\multicolumn{1}{c}{Present} &
\multicolumn{1}{c}{RCI \cite{om}}\\
\hline
\endfirsthead\\
\caption[]{(continued)}
Index &
\multicolumn{1}{c}{Configuration} &
\multicolumn{1}{c}{Level} &
\multicolumn{1}{c}{Present} &
\multicolumn{1}{c}{RCI \cite{om}}\\
\hline
\endhead
80 & 3d$^3(^4_3$F$)$4p & $^5$F$_1^{\rm o}  $ &  0.549 & 0.457 \\
82 & 3d$^3(^4_3$F$)$4p & $^5$D$_1^{\rm o}  $ &  1.220 & 1.227 \\
89 & 3d$^3(^4_3$F$)$4p & $^5$F$^{\rm o}_1  $ &  0.231 & 0.317 \\
97 & 3d$^3(^4_3$P$)$4p & $^5$P$_1^{\rm o}  $ &  2.477 & 2.474 \\
101& 3d$^3(^4_3$P$)$4p & $^5$D$_1^{\rm o}  $ &  1.500 & 1.494 \\
104& 3d$^3(^4_3$P$)$4p & $^5$D$_1^{\rm o}  $ &  1.521 & 1.513 \\
122& 3d$^3(^2_3$P$)$4p & $^3$P$_1^{\rm o}  $ &  1.485 & 1.453 \\
126& 3d$^3(^2_3$P$)$4p & $^3$D$_1^{\rm o}  $ &  0.533 & 0.547 \\
132& 3d$^3(^2_3$D$)$4p & $^1$P$_1^{\rm o}  $ &  1.159 & 0.949 \\
133& 3d$^3(^2_3$P$)$4p & $^3$S$_1^{\rm o}  $ &  1.746 & 1.742 \\
138& 3d$^3(^4_3$P$)$4p & $^3$D$_1^{\rm o}  $ &  0.565 & 0.820 \\
144& 3d$^3(^2_3$D$)$4p & $^3$D$_1^{\rm o}  $ &  0.566 & 0.536 \\
150& 3d$^3(^2_3$D$)$4p & $^3$P$_1^{\rm o}  $ &  1.448 & 1.485 \\
156& 3d$^3(^4_3$P$)$4p & $^3$S$_1^{\rm o}  $ &  1.998 & 1.915 \\
157& 3d$^3(^2_3$P$)$4p & $^1$P$_1^{\rm o}  $ &  1.003 & 1.063 \\
168& 3d$^3(^2_3$F$)$4p & $^3$D$_1^{\rm o}  $ &  0.500 & 0.500 \\
171& 3d$^3(^2_1$D$)$4p & $^3$D$_1^{\rm o}  $ &  0.506 & 0.509 \\
179& 3d$^3(^2_1$D$)$4p & $^3$P$_1^{\rm o}  $ &  1.494 & 1.490 \\
182& 3d$^3(^2_1$D$)$4p & $^1$P$_1^{\rm o}  $ &  1.000 & 0.999 \\
\hline   
\end{longtable}

\clearpage


\begin{longtable}{rrrrrrr}
\caption{\label{table_D}
Comparison of oscillator strengths ($f$-values) for some E1 transitions of Fe~V.
}
$i$ &
\multicolumn{1}{c}{$j$} &
\multicolumn{1}{c}{SE} &
\multicolumn{1}{c}{RCI$_{\rm L}$} &
\multicolumn{1}{c}{RCI$_{\rm V}$} &
\multicolumn{1}{c}{BPRM} &
\multicolumn{1}{c}{QR} \\
\hline
\endfirsthead
\caption[]{(continued)}
$i$ &
\multicolumn{1}{c}{$j$} &
\multicolumn{1}{c}{HRF} &
\multicolumn{1}{c}{RCI$_{\rm L}$} &
\multicolumn{1}{c}{RCI$_{\rm V}$} &
\multicolumn{1}{c}{BPRM}  &
\multicolumn{1}{c}{QR}\\
\hline
\endhead
  1  &  80  &  0.163  &  0.110  &  0.116  &  0.2154  &  0.1366 \\
  1  &  82  &  0.041  &  0.060  &  0.064  &  0.0055  &  0.0702 \\
  1  &  89  &  0.059  &  0.061  &  0.065  &  0.0574  &  0.0321 \\
  1  &  97  &  0.076  &  0.072  &  0.073  &  0.0842  &  0.0755 \\
  6  &  80  &  0.039  &  0.036  &  0.041  &  0.0231  &  0.0271 \\
  6  &  89  &  0.061  &  0.046  &  0.051  &  0.0670  &  0.0410 \\
  6  & 122  &  0.153  &  0.141  &  0.148  &  0.0938  &  0.1360 \\
  6  & 133  &  0.028  &  0.011  &  0.012  &  0.0022  &  0.0277 \\
  6  & 144  &  0.024  &  0.020  &  0.020  &  0.0071  &  0.0385 \\
 23  & 122  &  0.010  &  0.010  &  0.011  &  0.0070  &         \\
 23  & 132  &  0.216  &  0.108  &  0.118  &  0.0080  &  0.1560 \\
 23  & 133  &  0.010  &  0.042  &  0.045  &  0.0002  &  0.0277 \\
 23  & 138  &  0.029  &  0.054  &  0.059  &  0.0020  &  0.0257 \\
 23  & 150  &  0.011  &  0.012  &  0.013  &  0.0115  &         \\
 23  & 157  &  0.073  &  0.059  &  0.060  &  0.0786  &  0.0729 \\
 31  & 104  &  0.013  &  0.010  &  0.011  &  0.0101  &  0.0076 \\
 31  & 150  &  0.021  &  0.016  &  0.017  &  0.0520  &         \\
 31  & 156  &  0.088  &  0.074  &  0.082  &  0.0482  &  0.0665 \\
 31  & 168  &  0.168  &  0.136  &  0.145  &  0.1648  &  0.1390 \\
 31  & 179  &  0.046  &  0.045  &  0.042  &  0.0487  &  0.0429 \\
 34  & 182  &  0.379  &  0.289  &  0.295  &  0.3468  &  0.2850 \\
\hline									 	
\end{longtable}

\begin{flushleft}
SE: Calculations of Fawcett \cite{bcf} with the semi-empirical relativistic atomic structure code  \\
RCI$_{\rm L}$: Calculations of O`Malley et al. \cite{om} in the length form with the RCI code \\
RCI$_{\rm V}$: Calculations of O`Malley et al. \cite{om} in the velocity form with the RCI code \\
BPRM: Calculations of Nahar and Pradhan\cite{np} with the BPRM code \\
QR: Present calculations with the QR code \\
\end{flushleft}

\clearpage


\begin{longtable}{rrrrrrrrrr}
\caption{\label{table_E}
Comparison of  radiative rates ($A$-values, s$^{-1}$) for some M1 transitions 
among the levels of the 3d$^4$ configuration of Fe~V. \\
$a{\pm}b \equiv a{\times}$10$^{{\pm}b}$.
}
$i$ &
\multicolumn{1}{c}{$j$} &
\multicolumn{1}{c}{Garstang \cite{gar}} &
\multicolumn{1}{c}{Present}  &
$i$ &
\multicolumn{1}{c}{$j$} &
\multicolumn{1}{c}{Garstang \cite{gar}} &
\multicolumn{1}{c}{Present} \\
\hline
\endfirsthead
\caption[]{(continued)}
$i$ &
\multicolumn{1}{c}{$j$} &
\multicolumn{1}{c}{Garstang \cite{gar}} &
\multicolumn{1}{c}{Present}  &
$i$ &
\multicolumn{1}{c}{$j$} &
\multicolumn{1}{c}{Garstang \cite{gar}} &
\multicolumn{1}{c}{Present} \\
\hline
\endhead
  1  &    2  &  1.6$-$4  &  1.59$-$4  &    7  &    9  &  6.5$-$4  &  6.06$-$4 \\
  1  &    8  &  1.3$-$1  &  1.38$-$1  &    7  &   15  &  3.6$-$2  &  4.49$-$2 \\
  1  &   21  &  2.2$-$1  &  2.62$-$1  &    7  &   16  &  3.3$-$2  &  3.60$-$2 \\
  2  &    3  &  1.2$-$3  &  1.16$-$3  &    7  &   19  &  1.8$-$1  &  1.81$-$1 \\  
  2  &    6  &  1.3$-$0  &  1.62$-$0  &    8  &   21  &  1.2$-$1  &  1.39$-$1 \\
  2  &   12  &  1.0$-$1  &  1.18$-$1  &    8  &   24  &  6.2$-$2  &  8.23$-$2 \\
  2  &   20  &  2.0$-$1  &  2.18$-$1  &    9  &   10  &  5.8$-$4  &  5.68$-$4 \\
  2  &   21  &  1.9$-$1  &  2.32$-$1  &    9  &   17  &  4.1$-$2  &  4.95$-$2 \\
  3  &    4  &  2.6$-$3  &  2.64$-$3  &    9  &   19  &  2.5$-$1  &  2.69$-$1 \\
  3  &    8  &  1.1$-$0  &  1.23$-$0  &    9  &   22  &  1.1$-$1  &  1.29$-$1 \\
  3  &   12  &  2.0$-$1  &  2.31$-$1  &   10  &   17  &  4.1$-$2  &  4.98$-$2 \\
  3  &   13  &  1.6$-$1  &  2.02$-$1  &   10  &   22  &  1.4$-$1  &  1.70$-$1 \\
  3  &   15  &  7.0$-$3  &  8.10$-$3  &   11  &   18  &  5.6$-$2  &  6.21$-$2 \\
  3  &   18  &  9.7$-$2  &  1.08$-$1  &   11  &   20  &  5.2$-$2  &  6.13$-$2 \\
  3  &   20  &  1.8$-$1  &  1.76$-$1  &   11  &   21  &  3.6$-$2  &  3.78$-$2 \\
  4  &    5  &  3.0$-$3  &  2.91$-$3  &   11  &   24  &  1.8$-$1  &  2.26$-$1 \\
  4  &    7  &  4.0$-$4  &  1.18$-$3  &   12  &   15  &  3.0$-$2  &  3.32$-$2 \\
  4  &   11  &  7.1$-$1  &  7.93$-$1  &   12  &   24  &  2.1$-$1  &  2.64$-$1 \\
  4  &   12  &  4.7$-$2  &  6.39$-$2  &   13  &   15  &  3.7$-$2  &  4.14$-$2 \\
  4  &   13  &  4.0$-$1  &  5.30$-$1  &   13  &   19  &  1.5$-$1  &  1.49$-$1 \\
  4  &   14  &  1.6$-$1  &  1.87$-$1  &   13  &   24  &  4.2$-$1  &  5.09$-$1 \\
  4  &   15  &  1.7$-$2  &  1.94$-$2  &   14  &   16  &  2.7$-$2  &  3.04$-$2 \\
  4  &   16  &  7.8$-$2  &  8.59$-$3  &   14  &   17  &  3.7$-$2  &  4.01$-$2 \\
  4  &   18  &  8.9$-$2  &  1.01$-$1  &   14  &   19  &  3.2$-$1  &  3.44$-$1 \\
  4  &   20  &  1.1$-$1  &  1.20$-$1  &   15  &   19  &  4.2$-$2  &  4.59$-$2 \\
  5  &    7  &  1.1$-$3  &  6.24$-$3  &   15  &   25  &  1.2$-$1  &  1.35$-$1 \\
  5  &   13  &  6.6$-$2  &  7.21$-$2  &   16  &   25  &  1.7$-$1  &  1.86$-$1 \\
  5  &   14  &  7.4$-$1  &  8.98$-$1  &   18  &   24  &  9.0$-$2  &  1.19$-$1 \\
  5  &   16  &  3.2$-$2  &  3.02$-$2  &   18  &   25  &  1.5$-$1  &  1.79$-$1 \\
  5  &   18  &  3.7$-$1  &  4.20$-$1  &   20  &   25  &  7.0$-$2  &  7.93$-$2 \\
  6  &   21  &  4.9$-$2  &  5.91$-$2  &   21  &   24  &  8.0$-$2  &  1.05$-$1 \\
\hline									 	
\end{longtable}

\clearpage


\begin{longtable}{rrll}
\caption{\label{table_F}
Comparison of radiative rates ($A$-values, s$^{-1}$) for some E2 transitions 
of Fe~V. $a{\pm}b \equiv a{\times}$10$^{{\pm}b}$.
}
$i$ &
\multicolumn{1}{c}{$j$} &
\multicolumn{1}{c}{BPRM \cite{sn}} &
\multicolumn{1}{c}{Present} \\
\hline
\endfirsthead
\caption[]{(continued)}
$i$ &
\multicolumn{1}{c}{$j$} &
\multicolumn{1}{c}{BPRM \cite{sn}} &
\multicolumn{1}{c}{Present} \\
\hline
\endhead
  1 & 36 & 7.79+3 & 4.19+3 \\
  2 & 35 & 1.54+4 & 8.27+3 \\
  2 & 37 & 1.01+4 & 5.42+3 \\
  3 & 35 & 1.09+4 & 5.87+3 \\
  3 & 36 & 1.26+4 & 6.74+3 \\
  3 & 38 & 1.01+4 & 5.40+3 \\
  4 & 36 & 6.83+3 & 3.66+3 \\
  4 & 37 & 1.35+4 & 7.23+3 \\
  4 & 38 & 6.97+3 & 3.74+3 \\
  4 & 39 & 7.09+3 & 3.81+3 \\
  5 & 37 & 2.94+3 & 1.57+3 \\
  5 & 38 & 1.09+4 & 5.81+3 \\
  5 & 39 & 2.10+4 & 1.13+4 \\
\hline									 	
\end{longtable}

\end{document}